\documentclass[12pt,a4paper]{article}

\newcommand{\be}{\begin{equation}}
\newcommand{\ee}{\end{equation}}
\newcommand{\bea}{\begin{eqnarray}}
\newcommand{\eea}{\end{eqnarray}}

\newcommand{\Tr}{\mathop{\mathrm{Tr}}\nolimits}
\newcommand{\dd}{\mathrm{d}}
\newcommand{\ket}[1]{\left|#1\right\rangle}
\newcommand{\bra}[1]{\left\langle #1\right|}

\begin{document}

\title{Noncommutative radial waves}

\author{Ciprian Acatrinei\thanks{On leave from: {\it Institute of
        Atomic Physics  -
        P.O. Box MG-6, 76900 Bucharest, Romania}; e-mail:
        acatrine@physics.uoc.gr.} \\
        Department of Physics, University of Crete, \\
        P.O. Box 2208, Heraklion, Greece}

\date{30 May, 2001}

\maketitle

\begin{abstract}
We study radial waves in (2+1)-dimensional noncommutative
scalar field theory, using operatorial methods.
The waves propagate along a discrete radial coordinate
and are described by finite series deformations of Bessel-type functions.  
At radius much larger than the noncommutativity scale $\sqrt{\theta}$, 
one recovers the usual commutative behaviour.
At small distances, classical divergences are smoothed out by noncommutativity. 
\end{abstract}

Field theories defined over a noncommutative space \cite{nc1} are interesting,
nonlocal but most probably consistent, deformations of the usual ones.
They also arise as a particular low energy limit of string theory in a B-field 
\cite{st1,st2}.
Noncommutative (NC) field theories 
display an intriguing IR/UV mixing (see \cite{ir_uv} and later works), demonstrated 
perturbatively but expected to hold in general. At the classical level, they 
possess solitonic solutions 
\cite{ncsol}
which have no obvious counterpart in local field theory.
Other nontrivial solutions of the equations of motion were also found, cf. for instance
\cite{other}.
However, in spite of considerable progress, 
NC field theories are far from being well understood, even classically.
One interesting issue, which might shed further light on these theories,
is the description of oscillation and propagation processes on fuzzy spaces.
An example will be discussed here.

The aim of this note is to describe radial waves 
in a $(2+1)$-dimensional free scalar field theory
with NC spatial coordinates 
(theories with NC time are believed to be non-unitary \cite{st}).
In contrast to plane waves, radial waves are affected by the presence of noncommutativity.
They propagate on a discrete space, 
provided by the eigenvalues of the radius square operator.
Their amplitude solves a discrete wave equation
and is given by a finite series, reminiscent of Bessel-type functions.
In the large radius limit 
(analogous to the `large quantum number limit' in quantum mechanics)
the number of terms of the series grows indefinitely.
Then, NC radial waves behave like the usual commutative ones,
being described by the asymptotics of cylindrical functions. 
At small radius, 
the solutions of the wave equation are nonsingular, as they are everywhere,
even in the presence of sources.
Thus NC field theories may offer a solution to the old problem of 
{\it classical} divergences, although they do not seem to qualify
at the quantum level.

We will use operatorial methods, 
which are quite straightforward in the present context.
The Weyl-Moyal approach would require an unambiguous way
to switch between Cartesian and polar NC coordinates.

{\bf The set-up}

Let us start with the following action, written in operatorial form, 
\be
S=\int\dd t \Tr_{{\cal H}} 
\left (
\frac{1}{2}(\dot{\Phi}^2+\frac{1}{2}[X_i, \Phi]^2 
\right ), \qquad i=1,2.
\ee
The scalar field $\Phi$ is a time-dependent operator 
acting on the Hilbert space ${\cal H}$ on which the algebra
\be
[x_1,x_2]=i\theta \label{cr}
\ee
is represented. 
There is no potential term, $V(\Phi)=0$, since we study free waves.
$X_{i}$ is given 
by $X_{i}=p_{i}+A_{i}$, where 
$p_i=\theta^{-1}\epsilon_{ij}x^{j}$.
In the following the gauge field  $A_{i}$ is taken to be zero;
consequently we dropped further parts of the action which depend on it.
The equations of motion for the field $\Phi$ are 
\be
\ddot{\Phi}+[X_{i},[X_{i},\Phi]]=\ddot{\Phi}+\frac{1}{\theta^2}[x_i,[x_i,\Phi]]=0. \label{em}
\ee
In Cartesian coordinates, the solution of (\ref{em}) is straightforward,
\be
\Phi \sim e^{i(k_1 x_1 + k_2 x_2)-i\omega t}, \quad k_1^2+k_2^2=\omega^2, \label{plane}
\ee
and describes plane waves, which are formally identical to the commutative ones.
(However (\ref{plane}) has in fact bilocal character, see e.g. \cite{acatrinei}). 

The novelty appears when one considers polar coordinates.
If one chooses the oscillator basis $\{\ket{n}\}$ given by
\be
N\ket{n}=n\ket{n}, \quad N=\bar{a}a, \quad
a=\frac{1}{\sqrt{2\theta}}(x_1+ix_2), 
\ee 
the equations of motion become
\be
\ddot{\Phi}+\frac{2}{\theta}[a,[\bar{a},\Phi]]=0. \label{em2}
\ee
$N=\frac{1}{2}(\frac{x_1^2+x_2^2}{\theta}-1)$ 
is basically the radius square operator, in units of $\theta$.
Thus, radial symmetry amounts to the assumption $\Phi =\Phi (N)$.
Then  $\Phi$ is diagonal in the $\ket{n}$ basis, and its components  
are time-dependent c-numbers, $\bra{n}\Phi (t)\ket{n}\equiv \Phi_n (t)$. 
They satisfy the equation
\be
\ddot{\Phi}_n-\frac{2}{\theta}(n\Delta^2 \Phi_{n-1}+\Delta\Phi_n)=0,
\quad n=0,1,2,\dots         \label{em3}
\ee
in which the discrete derivative operator $\Delta$ is defined by
\be
\Delta (\Phi_n)=\Phi_{n+1}-\Phi_n.
\ee
(Alternatively, one could have diagonalized $\Phi$ in the $\ket{n}$ basis,
without assuming radial symmetry. This can be done only once however,
thus only one particular solution of (\ref{em2}) would be diagonal.
The others would be bilocal, i.e. of the form $\bra{n'}\Phi (t)\ket{n}$,
with $n'\neq n$.) 

If one assumes the time dependence of $\Phi_n$ to be of the form 
$e^{i\omega t}$,
one gets the difference equation
\be
n\Delta^2 \Phi_{n-1}+\Delta\Phi_n+\lambda\Phi_n=0, 
\quad n=0,1,2,\dots        \label{em4}
\ee
where $\lambda=\theta\omega^2 /2$ for a massless scalar field.
($2\lambda/\theta =\omega^2-m^2$ for a massive field, 
and $2\lambda /\theta=\omega^2+m^2$ for a tachyon.)

{\bf Solution of the equation of motion}

Equation (\ref{em4}) describes travelling or stationary waves on a 
semi-infinite discrete space,
namely the points $n=0,1,2,\dots$. 
We will find its two independent linear solutions
in the form of (eventually finite) power series.
To do so, notice that the standard power $n^k=n\cdot n\cdot\dots\cdot n$ 
does not behave simply under the action of $\Delta$. 
To adapt the usual logic of power series solutions to the above discrete equation,
we define a different type of `power of $n$' ($n$ is still a positive integer): 
\be
n^{(k)}=n(n-1)(n-2)\dots(n-k+1)=\frac{n!}{(n-k)!}.
\ee
It has the quite useful property that $\Delta n^{(k)}=k n^{(k-1)}$.

We now try to find a solution $\Phi(n,\sigma)$ of the form
\be
\Phi(n,\sigma)=\sum_{k=0}^{\infty}a_k(\sigma) n^{(k+\sigma)}, \label{phi}
\ee
with $\sigma$ an arbitrary parameter, to be fixed by the equation.
Substituting (\ref{phi}) into (\ref{em4}), one obtains a recurrence relation 
for the coefficients $a_k(\sigma)$,
\be
a_k(\sigma)=\frac{(-\lambda)}{(k+\sigma)^2}a_{k-1}(\sigma)
=\frac{(-\lambda)^k}{(k+\sigma)^2(k-1+\sigma)^2\dots(1+\sigma)^2}a_0 \label{a_k}
\ee
and a condition for $\sigma$,
\be
\sigma^2=0, \quad \mbox{i.e.} \quad \sigma_1=\sigma_2=0. \label{index}
\ee
Thus a first solution of our equation is
\be
\Phi_1(n)=a_0\sum_{k=0}^{n} \frac{(-\lambda)^k}{(k!)^2} n^{(k)}. \label{phi1}
\ee
It is given by a finite sum, since $n^{(p>n)}=0$ for $n$, $p$  positive integers.
It is understood that,
if one calculates a discrete derivative of (i.e. apply $\Delta$ to) the above solution,
one should put $\sigma=0$ only after operating with $\Delta$. 
$a_0$ is a dimensionful constant, which will be dropped from now on; 
it can be reinstated at any moment.
 
Since equation (\ref{index}) has two equal roots,
the above procedure provides only one solution of (\ref{em4}). 
Adapting again the methods used for continuous variables 
(see for instance \cite{BdiP}) 
to the discrete case,
a second linearly independent solution of (\ref{em4}) can be shown to be given by
\be
\Phi_2(n)=\left [ \frac{\partial \Phi(n,\sigma)}{\partial \sigma} \right ]_{\sigma=0}.
\ee
In the above equation, $\Phi(n,\sigma)$ has the form (\ref{phi}),
with $a_k(\sigma)$ given by (\ref{a_k}).
In order to actually evaluate $\Phi_2(n)$, 
we need to take the derivatives with respect to $\sigma$
of $a_k(\sigma)$, and of $n^{(k+\sigma)}$, at $\sigma =0$. 
It is easy to show, using (\ref{a_k}), that
\be
\left.\frac{d}{d\sigma}a_k(\sigma)\right |_{\sigma = 0}=
-2 a_k(0) H_k,
\ee
where $H_k$ is (up to a constant) a discrete version of the logarithmic function,
\be
H_k=1+\frac{1}{2}+\frac{1}{3}+\dots +\frac{1}{k}.\label{H_k}
\ee
To find $\frac{d}{d\sigma} n^{(k+\sigma)}$,
we have to extend the definition of $n^{(k+\sigma)}$ to arbitrary real powers.
Having in mind the representation of the factorial
by Gamma functions, $n!=\Gamma(n)=\int_{0}^{\infty}dt e^{-t} t^n$,
we define $n^{(k+\sigma)}=\frac{\Gamma(n)}{\Gamma(n-k-\sigma)}$. 
In the end, one obtains,
dropping the part of the result which is proportional to $\Phi_1(n)$,
\be
\Phi_2(n)=\sum_{k=0}^{n-1} \frac{(-\lambda)^k}{(k!)^2} n^{(k)}  (H_{n-k}-2 H_{k}). 
\label{phi2}
\ee
The general solution of (\ref{em4}) is a linear combination of 
$\Phi_1(n)$ and $\Phi_2(n)$,
\be
\Phi(n)=c_1\Phi_1(n)+c_2\Phi_2(n), \label{gensol}
\ee 
with the coefficients $c_{1,2}$ fixed by some given boundary conditions.

{\bf Large distances: commutative limit}

We now consider the $n\rightarrow\infty$ limit, in order to see how the
precedent solutions behave at distances $r>>\sqrt{\theta}$ from the $n=0$ point.
(This can also be seen as a small $\theta$ limit.)
Using $\lambda =\theta \omega^2 /2$ and 
$ n=\frac{r^2}{2\theta}\rightarrow\infty$, $\Phi_1(n)$ becomes,
as a function of $r$, the zero-order Bessel function of first type:
\be
\Phi_1(n) 
\stackrel{n\rightarrow\infty}{\rightarrow} 
f_1(r)=\sum_{k=0}^{\infty}
\frac{(-1)^k(\omega r)^{2k}}{(k!)^2 2^{2k}}=J_0(\omega r)
\stackrel{r\rightarrow\infty}{\sim}
\sqrt{\frac{2}{\pi \omega r}}cos(\omega r-\pi /4) .\label{j}
\ee
We see that $f_1(r)$ is independent of $\theta$.
This is not the case for the function of $r$ 
one would obtain at finite $n$, which diverges as $\theta\rightarrow 0$.
We will encounter only Bessel functions of zero order in what follows, 
since the angular dependence of $\Phi$ is lost, due to (\ref{cr}).
We stress that the absence of radial symmetry would not introduce
angular dependence in the solutions of the equations of motion.
It would just lead to nonzero (bilocal) solutions of the type
$\bra{n'} \Phi \ket{n}, n'\neq n$.

Similarly, $\Phi_2$ becomes
\be
\Phi_2(n) \rightarrow f_2(r)=\sum_{k=0}^{\infty}
\frac{(-1)^k(\omega r)^{2k}}{(k!)^2 2^{2k}}
[2ln(\omega r)-2H_k+\gamma-ln(2\theta\omega^2)]. \label{f2}
\ee
$\gamma$ is the Euler-Mascheroni constant, 
$\gamma=\lim_{k=\infty}(H_k-ln k) \simeq 0.5772.$
Thus $f_2(r)$ still depends on $\theta$, via a logarithmic term,
which renders the  $\theta\rightarrow 0$ limit singular.
Using the series expansion of
the Bessel function of second kind (Neumann function) \cite{BdiP}, 
$Y_0(\omega r)
\stackrel{r\rightarrow\infty}{\sim} 
\sqrt{\frac{2}{\pi \omega r}}sin(\omega r-\pi /4)$,
\be
Y_0(\omega r)=\frac{2}{\pi}
\left ( 
\sum_{k=0}^{\infty}
\frac{(-1)^{k+1}(\omega r)^{2k}}{(k!)^2 2^{2k}}H_k
+(\gamma+ln(\omega r/2))J_0(\omega r)
\right )
\ee
one sees that $f_2(r)$ is a linear combination of Bessel functions of first 
and second kind: 
$f_2(r)/\pi= Y_0(\omega r)+(\gamma+ln(2\theta\omega^2))J_0(\omega r)$.

Hence, the $n\rightarrow\infty$ limits of the solutions $\Phi_1(n)$ and
$\Phi_2(n)$ obey the Bessel equation. This is in agreement with the 
$n=\frac{r^2}{2\theta}\rightarrow\infty$ 
limit of the difference operator entering equation (\ref{em4}), 
\be
\frac{2}{\theta}(n\Delta^2 \Phi_{n-1}+\Delta \Phi_n) 
\stackrel{n\rightarrow\infty}{\rightarrow}
\frac{2}{\theta}(n\frac{d^2}{d n^2}+\frac{d}{d n})\Phi(n)
\stackrel{n=\frac{r^2}{2\theta}}{=}
(\frac{d^2}{d r^2}+\frac{1}{r}\frac{d}{d r})f(r).
\ee
Thus, at $r>>\sqrt{\theta}$, NC radial waves behave like
ordinary, comutative ones.
This allows us to find, via a 'correspondence principle', 
which combinations of $\Phi_1(n)$ and $\Phi_2(n)$ correspond to stationary,
respectively travelling NC waves: 
In the commutative case, 
standing waves are described by the $J_0(r)$ function,
whereas radially expanding ones
by the first Hankel function $H^{(1)}_{0}(r)=J_0(r)+ i Y_0(r)$.
Hence, the linear combination of $\Phi_1(n)$ and $\Phi_2(n)$
which at $n\rightarrow\infty$ tends to $J_0(\omega r)$
will describe standing noncommutative waves. 
This is obviously $\Phi_1(n)$,
which consequently solves finite area boundary value problems with radial symmetry,
describing standing oscillations.
On the other hand, the function which tends to $H_0(\omega r)$ as $r\rightarrow\infty$,
namely 
\be
\Phi_3(n)=\Phi_1(n)+\frac{i}{\pi}
\left ( \Phi_2(n)+[\gamma+ln(\theta\omega^2 /2)]\Phi_1(n)
\right ) ,
\label{phi3}
\ee
represents a radial NC wave propagating outwards, towards $n=\infty$. 
Any solution $\Phi(n)$ of (\ref{em4}) 
can be written as a linear superposition of $\Phi_1(n)$ and 
either $\Phi_2(n)$ or $\Phi_3(n)$,
with coefficients determined by the boundary conditions one wishes to impose.
It is understood that all the above solutions are multiplied by a dimensionful,
otherwise arbitrary, constant; the same will apply for sources.

{\bf Small distance: no classical divergences}

It is worth noting that, in sharp contrast to the commutative case, 
in which Hankel and Neumann functions are singular at the origin,
the functions $\Phi_{2,3}$ are nowhere singular (except when $\theta=0$).
This suggests that, although not finite in quantum perturbation theory, 
fields defined over noncommutative spaces
may not display {\it classical} divergences. 
This happens simply because the sources are not localized
(also, one has no access to the origin: $r / \sqrt{\theta}=\sqrt{2n+1}\geq 1$).
In order to rigorously support such a claim, one should include sources in the
calculation, i.e, solve the inhomogeneous version of equation (\ref{em4}).
We will present here a strategy to achieve this, 
and will demonstrate the non-divergent character of the solutions. 
Consider first a nonzero source $j(n=0)$, oscillating in time with frequency $\omega$.
It modifies equation (\ref{em4}) only at $n=0$, where one has
\be
\Delta\Phi(0)+\lambda\Phi(0)=\Phi(1)-(1-\lambda)\Phi(0)=j(0).
\ee 
The relation between $\Phi(0)$ and $\Phi(1)$ is then modified. 
This can be simply accomodated through
a change of the coefficients $c_1$ and $c_2$ in (\ref{gensol}),
i.e. a change of the respective weight of each independent solution.
Addition of a source at $n=0$ thus preserves the finite character of the solutions. 
The same applies for a source $j(n)$ added at an arbitrary single point $n$.
The change in the relation between $\Phi(n-1)$, $\Phi(n)$ and $\Phi(n+1)$ due to $j(n)$,
\be
(n+1)\Phi(n+1)-(2n+1-\lambda)\Phi(n)+n\Phi(n-1)=j(n)
\ee
is again easily taken int account by an appropriate change of the weights
in the homogeneous solution (\ref{gensol}).
(Imposition of nonzero $j(n)$ at some $n$ is equivalent to a specific boundary condition.)
The solution for an arbitrary distribution of charges $j(n), \forall n$, 
is now obtained by linear superposition of the above type of solutions. 
It does not display singularities.   

\vskip 0.3cm

Let us conclude with a summary of what we have shown:
\begin{itemize}

\item On the NC plane defined by $[x^1,x^2]=i\theta$,
radial waves propagate on a discrete space, given by the eigenvalues
$r=\sqrt{2n+1}, n=0,1,2,\dots$ of the radius square operator. 
One has no access to the origin ($r=0$), as one would expect.
The amplitude of the waves is given by a finite series, whose number of
terms depends on the location at which the field amplitude is calculated:
at radius  $r=\sqrt{2n+1}\sqrt{\theta}$, one has $n+1$ terms in the series.

\item In the large radius limit, $r>>\sqrt{\theta}$, or
$n\rightarrow\infty$, the amplitudes become Bessel-type functions,
consequently the waves behave like commutative ones.

\item At small radius, if $\theta\neq 0$,
there are no signs of singularities appearing.
This drastic improvement in the behaviour of classical noncommutative theories
deserves to be further explored. 

\end{itemize}

\subsection*{Acknowledgements}
This work was supported by a MURST (Italy) fellowship.
I thank the HEP theory group of the University of Crete for hospitality,
and Kostas Anagnostopoulos, Elias Kiritsis and 
Corneliu Sochichiu for useful discussions. 

\begin{thebibliography}{1}

\bibitem{nc1}
H.S. Snyder, 
\newblock Phys. Rev. 71 (1947) 38; Phys. Rev. 72 (1947) 68.
\newblock

\bibitem{st1} 
A. Connes, M.R. Douglas and A. Schwarz, 
\newblock JHEP  9802 (1998) 003, hep-th/9711162;
M.R. Douglas and C. Hull, JHEP 9802 (1998) 008.

\bibitem{st2}
N. Seiberg and E. Witten,
\newblock JHEP 9909 (1999) 032, hep-th/9908142.
\newblock 

\bibitem{ir_uv}
S. Minwalla, N. Seiberg and M. Van Raamsdonk, 
\newblock JHEP 0003 (2000) 035, hep-th/0002186.
\newblock 


\bibitem{ncsol}
R. Gopakumar, S. Minwalla and A. Strominger,
\newblock JHEP 0005 (2000) 020, hep-th/0003160.
\newblock 


\bibitem{other}
J.A. Harvey, P. Kraus, F. Larsen and E. J. Martinec,
\newblock JHEP 07 (2000) 042, hep-th/0005031;
\newblock 
D.J. Gross and N.A. Nekrasov, \newblock JHEP 07 (2000) 034, hep-th/0005204; 
\newblock
A.P. Polychronakos, \newblock Phys. Lett. B495 (2000) 407,  hep-th/0007043; 
\newblock
D. Bak, \newblock Phys. Lett. B495 (2000) 251, hep-th/0008204. 
\newblock

\bibitem{st} J. Gomis and T. Mehen, Nucl.Phys.  B591 (2000) 265, hep-th/0005129;
L. Alvarez-Gaum\'{e}, J.L.F. Barb\'{o}n and R. Zwicky, JHEP 0105 (2001) 057,
hep-th/0103069.
See however
D. Bahns, S. Doplicher, K. Fredenhagen and G. Piacitelli,   
Phys.Lett. B533 (2002) 178, hep-th/0201222.


\bibitem{acatrinei} C. Acatrinei, hep-th/0204197; see also hep-th/0210040.
\bibitem{BdiP} 
W.E. Boyce and R.C. DiPrima, 
\newblock Elementary Differential Equations, Wiley \& Sons, 1996.
\newblock

\end{thebibliography}

\end{document}